\documentclass[iop]{emulateapj}

\usepackage{txfonts}
\usepackage{graphicx}
\usepackage{graphicx}
\usepackage{natbib}

\shorttitle{Transverse oscillations in chromospheric mottles}
\shortauthors{Kuridze et al.}

\begin{document}

\title{Transverse oscillations in chromospheric mottles}
\vskip1.0truecm

\author{D. Kuridze$^{1,4}$, R. J. Morton$^{2}$, R. Erd\'elyi$^{2}$, G. D. Dorrian$^{3}$, M. Mathioudakis$^{1}$, 
D. B. Jess${^1}$,  and F. P. Keenan${^1}$}
\affil{$^1$Astrophysics Research Centre, School of Mathematics and Physics, Queen's University, Belfast, BT7~1NN, Northern Ireland, UK;\\
e-mail: dkuridze01@qub.ac.uk}  
\affil{$^2$Solar Physics and Space Plasma Research Centre (SP$^2$RC), University of Sheffield, Hicks Building, Hounsfield Road, Sheffield S3 7RH, UK}
\affil{$^3$Institute of Astronomy and Astrophysics, National Observatory of Athens, Lofos Nymfon, Thiseio, P.O.Box 20048, GR-11810 Athens, GREECE}
\affil{$^4$Abastumani Astrophysical Observatory at Ilia State University, University Str. 2, Tbilisi, Georgia} 

\begin{abstract}
A number of recent investigations have revealed that transverse waves are ubiquitous in the solar chromosphere. 
The vast majority of these have been reported in limb spicules and active region fibrils.  We investigate long-lived, quiet Sun, on-disk features such as chromospheric mottles 
(jet-like features located at the boundaries of supergranular cells) and their transverse motions.
The observations were obtained with the Rapid Oscillations in the Solar Atmosphere (ROSA) instrument at the Dunn Solar Telescope.  
The dataset comprises  simultaneous imaging in the H$\alpha$ core, Ca II K, and G band of an on-disk quiet Sun region. 
Time-distance techniques are used to study the characteristics of the transverse oscillations.
We detect over 40 transverse oscillations in both bright and dark mottles, with periods ranging from  70 to 280 s, 
with the most frequent occurrence at $\sim$ 165 s. The velocity amplitudes and transverse displacements exhibit characteristics similar to limb spicules.
Neighbouring mottles oscillating in-phase are also observed.
The transverse oscillations of  individual mottles are interpreted in terms of magnetohydrodynamic kink
waves. Their estimated periods and damping times are consistent with phase mixing and resonant mode conversion.
\end{abstract}


\section{Introduction}

Dark and bright mottles are commonly observed in the quiescent solar chromosphere. 
They are small-scale jet-like features of relatively cool and dense material  located at the boundaries of supergranular cells, ejected from the lower 
chromosphere at speeds of about  $10-40~\mathrm{km\,s^{-1}}$ \citep{tsir1,roup1}.
A number of earlier studies suggest that bright and dark mottles correspond to phenomena  
in the lower and upper chromosphere, respectively  \citep{bray1}, 
while others believe that the difference in brightness indicates different parts of the same structure 
\citep{beck1,beck2,ban,braylo,ster2}.  
Mottles are often considered as the disk representation of chromospheric 
spicules \citep{hanst3,voort1}.
The exact nature of mottles remains the subject of an ongoing debate, with the majority of solar researchers 
agreeing that mottles and spicules are related, in the sense that they have similar temperatures, density profiles, widths, 
lengths and lifetimes \citep{tsirop2,zach2}.
Chromospheric small-scale jet-like structures, such as mottles, spicules and fibrils, play an important role in the mass balance of the solar atmosphere.  
It is estimated that  only a small percentage  of the mass outflow provided by mottles is sufficient to 
compensate for the coronal mass loss due to the solar wind \citep{tsir1}.
 
The highly dynamic photosphere can excite magnetohydrodynamic
(MHD) waves which can propagate into the chromosphere and corona \citep{erd1,erd2}. 
Spicular structures can act as conduits for transferring wave energy  from the lower to the upper parts of the solar atmosphere. 
Numerous observations of transverse motions in spicular structures have been reported in recent years
\citep{kukh,dep1,he,tavabi,verth1,okamo}.
\begin{figure*}[t]
\begin{center}
\includegraphics[width=18.0cm]{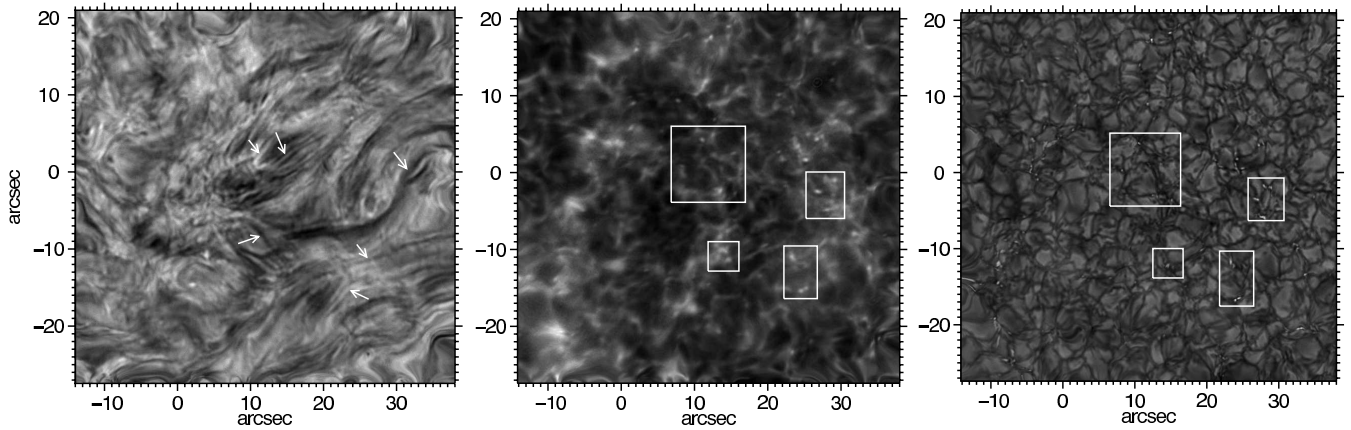}
\end{center}
\caption{Simultaneous ROSA images in H$\alpha$ core (left), Ca {\sc{ii}} K core (middle), and G band (right). 
Arrows indicate typical bright and dark mottles in H$\alpha$. 
White boxes indicate areas in which long-lived mottles  have been detected in H$\alpha$.  
Axes are in heliocentric  coordinates, where $1'' \approx 725$~km.}
\label{fig1}
\end{figure*}
An excellent review on the topic is given by \citet{zaq}. 
Kink waves have also been recently observed in active region  fibrils \citep{piet}.
These are often considered as on disk, active region spicules which connect areas of opposite 
magnetic polarity \citep{noort, dep1,gugl,kur1}.
Spicules and mottles are typically short-lived features, with lifetimes of the order of typical wave periods. The observed waves also appear to propagate with phase speeds close to the local Alfv\'en speed. 
Mottles and spicules have typical lengths of $\mathrm{4-6~Mm}$, so waves propagate along the mottle in only $\mathrm{60-90~s}$. 
Both of these features make it challenging to obtain information on the observed wave motion, e.g., amplitudes, periods, phase speeds.
However, the reported wave periods in spicules and fibrils are in the range  $\mathrm{1-5~mins}$. 
Transverse motions in spicules have, in the past, been misinterpreted as Alfven waves, yet spicules (and mottles) appear as overdense structures in observations that are assumed to outline the quiet Sun magnetic field.
In this case, the pure Alfv\'en waves are torsional motions  \citep{erdfed} and the transverse waves are 
the MHD kink modes.

Here we present high spatial and temporal resolution observations of transverse oscillations 
in on-disk bright and dark mottles  in the quiet solar chromosphere. 
We focus our attention on relatively long-lived mottles that display transverse 
motions which last for at least one wave cycle. We identify over 40 such waves and are able 
to obtain statistical information on the wave properties (e.g., amplitudes, periods).

\section{Observations and data reduction}
 
The observations were obtained between  15:41 - 16:51~UT on
29 September 2010 with the Rapid Oscillations in the Solar Atmosphere \citep{jess} 
imaging system, mounted on the Dunn Solar Telescope (DST) at the National Solar 
Observatory, New Mexico, USA. A typical dataset includes simultaneous imaging in G band, H$\alpha$ core, and
Ca {\sc{ii}} K. High-order adaptive optics were used throughout 
the observations. The images were reconstructed using the algorithms of  \citet{wog} followed by de-stretching. 
These algorithms were implemented to remove the effects of atmospheric distortion from the data. The effective cadence after 
reconstruction was 7.7~s for H$\alpha$,  9.6~s for Ca {\sc{ii}} K, and  1~s for  the G-band. The total field of view is  $48''\times52''$ 
with a spatial sampling  of $0.069''$/pixel  corresponding to a spatial resolution of  $\mathrm{150~km}$  in H$\alpha$.  

\section{Analysis and discussion}

Figure~\ref{fig1} shows co-spatial and co-temporal images of the field-of-view in the H$\alpha$ core, Ca II K and G-band.  
The H$\alpha$ image is dominated by elongated dark and bright mottles which appear co-spatial with Ca {\sc{ii}} K  
brightenings and photospheric magnetic bright points.  

\begin{figure*}[t]
\begin{center}
\includegraphics[width=17.9cm]{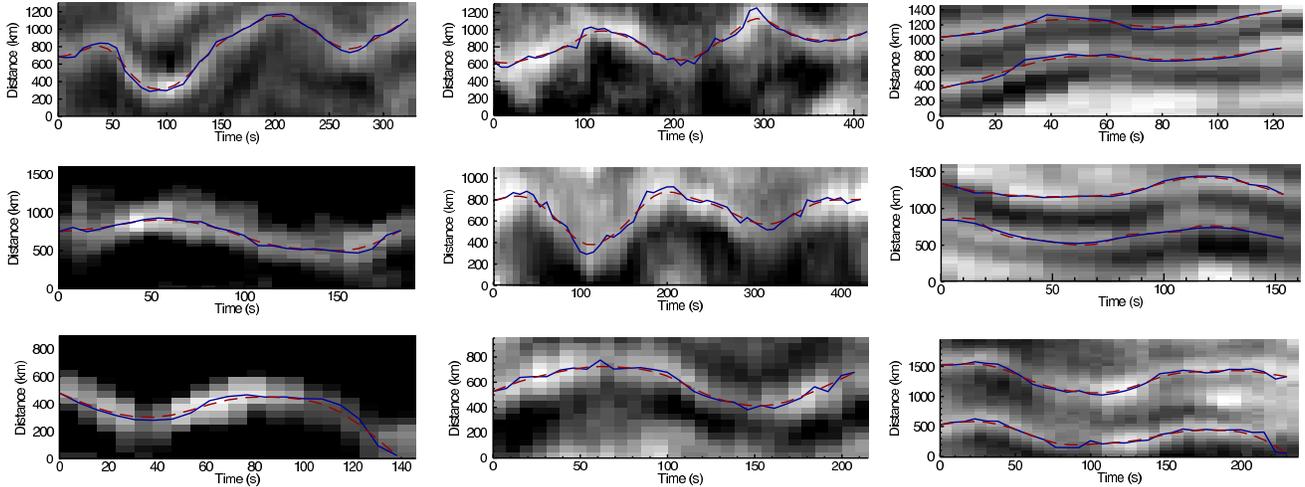}
\end{center}
\caption{Time-distance diagrams of the mottle intensity in H$\alpha$. The blue line is the centroid of a Gaussian fit to the cross-sectional flux profile of the mottle 
at each time-step  smoothed by a $\mathrm{\sim50~km}$ width (red line). }
\label{fig2}
\end{figure*}

We find a total of 42 mottles (23 bright, 19 dark)  that show transverse  waves lasting for at least 
one wave cycle. In 7 cases, neighbouring mottles are seen to oscillate in phase. 
Time-distance analysis reveals transverse oscillatory motions perpendicular to the mottle axis.
The centre is determined by fitting a Gaussian profile to the  cross-sectional flux profile for each time frame of  the transverse cut (blue lines in the Figure~\ref{fig2}). 
The overall motion of the mottle shows a linear trend which has been subtracted from the original signal. 
Oscillation periods were determined using wavelet analysis \citep{tor},   
and found to vary between $\mathrm{70-280~s}$, with a strong peak  at  $\sim\mathrm{165~s}$. 
Transverse velocities are in the range $\mathrm{3-18~km~s^{-1}}$, with displacement amplitudes between $\mathrm{100-400~km}$ (Figure~\ref{fig3}).  

The phase speed of the transverse motions can be evaluated by determining the phase difference 
between signals of the waves obtained at different heights along the mottle.
Phase difference analysis along the mottle axis can be undertaken accurately only for 7 mottles 
due to the complex fine structure of the H$\alpha$ images.  Estimated phase speeds are in the range  $\mathrm{40-110~km~s^{-1}}$. 
In most cases the oscillations are seen only in a segment of the mottle typically less than 1 Mm long. A cadence of $\sim$ 7.7 s 
limits the maximum phase speed that may be detected to $\sim$130 $\mathrm{km\,s^{-1}}$. The mean period, amplitude, and phase speed of the 
observed oscillations are similar to results found for limb spicules  \citep{kukh,zaq2,kim,he,okamo}.

\begin{figure*}[t]
\begin{center}
\includegraphics[width=15.9cm]{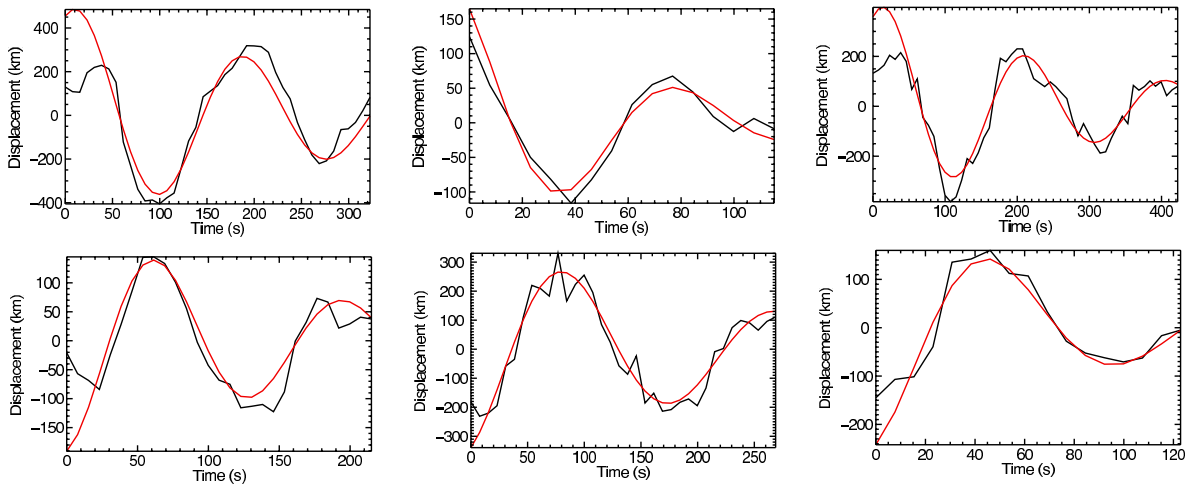}
\end{center}
\caption{The observed  mottle displacements (black line) fitted with an exponentially damped cosine function (red line).}
\label{fig4}
\end{figure*}

Mottles can be considered as straight, cylindrical, high density magnetic flux tubes which can guide different types of MHD wave modes, e. g., slow, fast, and  Alfv\'en.
Alfv\'en waves, propagating incompressibly along a magnetic flux tube, are torsional and  do not displace the tube axis.
MHD sausage modes may lead to intensity variations or variations of the mottle cross-section but  do not to cause a transverse displacement of the tube axis.
The clear transverse displacement of the mottle axis seen in our observations, suggests that the most appropriate interpretation 
for the oscillations is that of propagating/standing kink waves.
The characteristic speed for the transverse motions is the kink speed, $c_k$, 
defined as  \citep{edrob}: 

\begin{equation}
c^{2}_k={\rho_0v^{2}_{A0}+\rho_{e}v^{2}_{Ae}\over \rho_0+\rho_{e} },
\label{eq1}
\end{equation}
where $\rho_0$, $v_{A0}$, $\rho_e$ and $v_{Ae}$ are the plasma densities and Alfv\'en speeds inside and outside the tube, respectively.  
This equation is applicable near the thin tube limit, i. e. $r\ll\lambda$, where $\lambda$  and $r$ are the wavelength and mottle radius, respectively. 
The wavelength, $\lambda$, of the transverse wave is given by $\lambda=Tc_{k}$, 
where $T$ is the period of the wave. Hence $\lambda=\mathrm{4800~km}$ for $T=\mathrm{80~s}$ and $c_{k}=\mathrm{60~km~s^{-1}}$.
The radius of a mottle is typically $\sim200~\mathrm{km}$,  
hence  $r\ll\lambda$, the thin tube limit applies, and phase speed can be approximated by the kink speed. 
Furthermore, mottles are cool and dense material in the vicinity of strong magnetic flux concentration in network regions co-spatial 
with photospheric magnetic bright points (Figure~\ref{fig1}). They may therefore, be considered as low-$\beta$ plasma magnetic  tubes embedded in the field free (non magnetic) 
or weak field (compared to the mottle itself) environment. 
For no magnetic field outside the tube equation~(\ref{eq1}), implies  that  
\begin{equation}
c_k =\sqrt{{\rho_0\over \rho_0+\rho_e}}v_{A0}=\sqrt{{1\over 1+\rho_e/\rho_0}}{B_0\over\sqrt{4\pi\rho_0}}, 
\label{eq2}
\end{equation}
where $\mathrm{B_0}$ is the tube magnetic field strength. 
For a typical value of the tube magnetic field $B_{0}\approx10$ Gauss \citep{trujil}, density  $\rho_0\approx\mathrm{(1-3)}\times10^{-13}$ $\mathrm{g~cm^{-3}}$
\citep{beck2,ster,tsir1},
and  $\rho_e/\rho_0\sim0.1-0.033$
we estimate from equation~(\ref{eq2}) kink  speed range of $\sim\mathrm{46-86 ~km~s^{-1}}$. 
This value compares well with our measured phase speeds ($\mathrm{\sim40-110~km~s^{-1}}$).  
The observed periods show a strong peak at around 165 s (top right panel of the Figure~\ref{fig3}).  
This value is very close to the chromospheric kink wave cut-off period defined as  $T_{k}=8H\pi/c_{k}$, 
where $H$ is the chromospheric density scale height \citep{rae1,spru2,rob5}. 
For the kink speed discussed above and a chromospheric scale height of $H\sim\mathrm{500~km}$, 
we estimate a kink wave cut-off period of $T_k\sim\mathrm{170~s}$.
\begin{figure*}[t]
\begin{center}
\includegraphics[width=16.0cm]{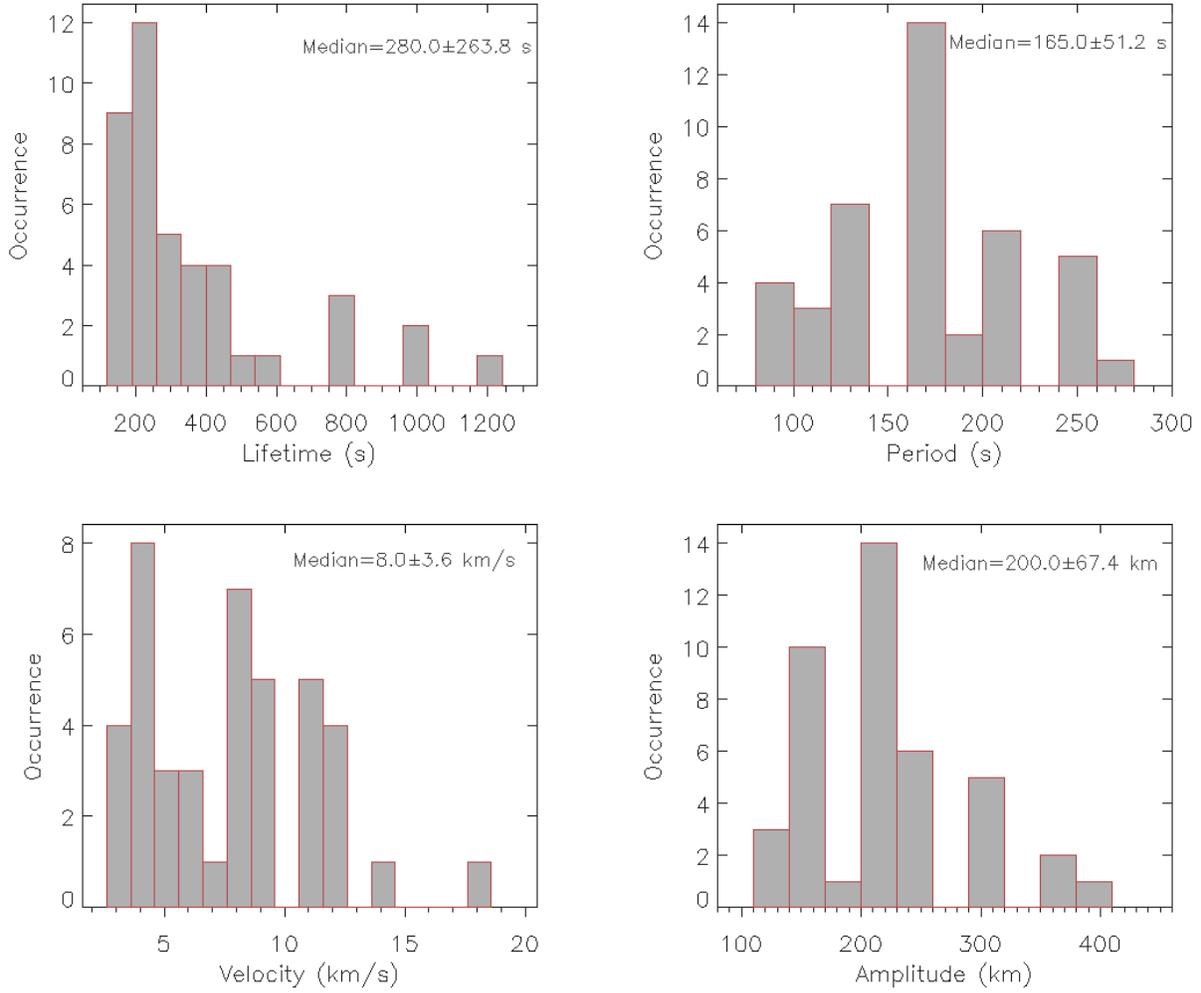}
\end{center}
\caption{The lifetime (top left),  period (top right), transverse velocity amplitude (bottom left),  
and transverse displacement amplitude (bottom right) for the 42 mottles analysed in this study.
Median values with standard deviations are also given.}
\label{fig3}
\end{figure*}
Waves with periods above the cut-off become evanescent. 
However, wave propagation in  non-ideal (e. g.  with ion-neutral collisions or non-adiabatic effects)
or inclined magnetic tubes can lead to an increase in the cut-off period and allows the penetration 
of lower-frequency oscillations higher up in the solar atmosphere  \citep{dep}.
Our imaging observations do not allow us to define accurate inclination angles for individual mottles. however, as  
we are looking at the projection of mottle lengths onto the plane of the sky, this implies 
that they are inclined from the vertical.
Detection  of periods above the cut-off value (Figure~\ref{fig3}) 
suggests that these waves may tunnel through the evanescent region by propagating across 
inclined magnetic field lines. 

The damping properties of the observed transverse oscillations  
are investigated by fitting the intensity profiles with an exponentially decaying cosine function, 
$z(t)=A\exp{(-t/\tau)}\cos(2\pi t/T+\phi)$, where the fitting parameters $A, T, \tau, \phi$ 
are the displacement amplitude, oscillation period and damping time, respectively \citep{asch}.
We have found 12 cases where the mottle oscillations show evidence for damping. 
In Figure~\ref{fig4} we highlight 6 examples fitted with the above function and the oscillation parameters determined are listed in Table~\ref{tabl1}. 
The time sequence is divided into 5 shorter overlapping sections, and errors are obtained from the maximum and the minimum decay rates and phase of the sections.

In Figure~\ref{fig5} we plot the estimated damping time as a function of
the oscillation period.  We also calculate a best-fit power law
scaling. The fit indicates an almost linear dependence on the
damping time of the kink wave period (Figure~\ref{fig5}).
Resonant mode conversion predicts that the damping time is a linear function of the period and may therefore be 
responsible for the damping of the oscillations \citep{goos1,ter1,soler} in the chromosphere. 
Phase mixing follows a scaling law of $\tau\sim T^{3/4}$ (Figure~\ref{fig5}) that can also be a plausible theory for the 
interpretation of the observed damping \citep{ofm2,mend}.
In contrast, a number of the transverse waves show an increase in amplitude with time. 
This suggests the possibility of a continuous driver or the influence of a time-dependent plasma 
\citep{mor,rud}. 
At present it is not possible to determine which, if any, of these options is responsible for the increasing amplitude.

\section{Concluding remarks}

\begin{table}[t]
\caption{The amplitude $A$,  oscillation period $T$,  damping time $\tau$ and phase $\phi$ for the damped oscillations.}
\begin{center}
\begin{tabular}{c   c   c   c   c}
\hline
\hline
Damping ~&~~A~~~&~~T~ ~&~~$\tau$~~~&~~$\phi$     \\ 
  cases   & (km)  & (sec)  & (sec) &  (rads)\\  
\hline    
\hline
1 & 400$\pm$75     &        165$\pm$12           &         285$\pm$70        &   0.45$\pm$0.15   \\ 
2 & 150$\pm$50    &         80$\pm$5               &         62$\pm$ 40          &   0.42$\pm$0.22\\ 
3 & 300$\pm$ 65     &       190$\pm$5      &                294$\pm$30            & 0.54$\pm$0.1\\                 
4 & 330$\pm$ 80     &       170$\pm$17       &             263$\pm$75               &   2.8$\pm$ 0.05  \\
5 & 170$\pm$ 60     &       120$\pm$12     &               191$\pm$35               &    3$\pm$ 0.5 \\
6 & 150$\pm$ 50     &     80$\pm$6       &                 266$\pm$ 100            &    3$\pm$ 0.1 \\
7 & 170$\pm$ 40     &     80$\pm$20       &               83$\pm$ 25             &    3$\pm$0.1 \\
8 & 250 $\pm$ 50    &       165 $\pm$16     &              476$\pm$ 100              &    2.9$\pm$ 0.1 \\
9 & 300$\pm$ 60     &    165$\pm$27       &              136$\pm$ 30              &    3.4$\pm$ 0.15 \\
10 & 250$\pm$ 70     &    165$\pm$26       &              200$\pm$ 65              &    0.12$\pm$ 0.05\\
11 & 150$\pm$60     &        170$\pm$14       &             147$\pm$ 65             &    0.11$\pm$0.04 \\
12 & 150$\pm$50     &     260$\pm$10       &              775$\pm$ 170              &    3.7$\pm$ 0.3\\
\hline
\hline
\end{tabular}
\end{center}
\label{tabl1}
\end{table}%

\begin{figure}[t]
\begin{center}
\includegraphics[width=8.3cm]{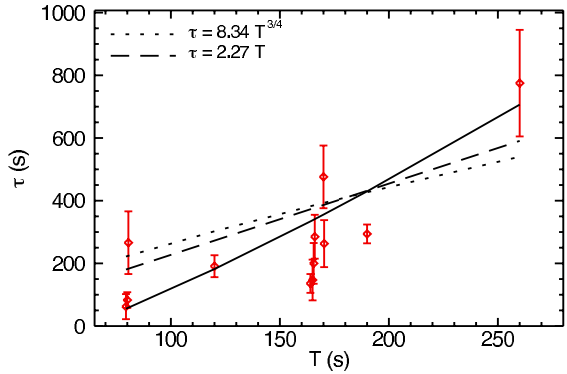}
\end{center}
\caption{Damping time $\tau$ vs period $T$ (red diamonds).   The solid line is the best fit scaling law. 
The dashed line is the fit to  $\tau=c_1T$ (scaling for resonant absorption) with $c_1=2.27$ 
and dotted line  is the fit to  $\tau=c_2T^{3/4}$ (scaling for phase mixing) with $c_2=8.34$.}  
\label{fig5}
\end{figure}

We have presented observations of relatively long-lived dark and bright mottles displaying transverse motions lasting for at least one wave cycle. 
The mottle oscillations studied in this work are interpreted as fast kink MHD waves. 
We obtain statistics on the properties, i.e. period, displacement and velocity amplitudes, of 42 separate events.
Coherent oscillations in neighbouring mottles indicate an oscillatory source that operates on spatial scales larger than the mottle itself.  
The distance between in-phase oscillating mottles varies between  $\mathrm{300-1000~km}$,  which is very similar to granular sizes. 
The relation between the damping times and periods suggests that resonant mode conversion and phase mixing may be viable damping mechanisms  (Figure~\ref{fig5}).
Dissipation of the wave energy associated with the transverse motions is thought to be 
important for the heating of the quiet solar corona and the acceleration of the solar wind \citep{cran}.

Possible excitation mechanisms for the waves include granular buffeting, global oscillations and reconnection events 
\citep{rob1,spru1,hol1,has1,dep,he2}. 
Recently,  \citet{jess2} suggested that 
longitudinal pressure modes in photospheric magnetic bright points 
can be converted  into the transverse oscillations observed in Type I spicules.
Following the successful approach of coronal seismology 
\citep{rud1,doors}, 
it may be possible that the observed kink waves could be exploited for  seismology of the chromosphere 
(e.g., Verth et~al. 2011). The higher temporal and spatial resolution of ground-based observing instruments 
(e.g., ROSA/DST, Crisp Imaging SpectroPolarimeter (CRISP)/Swedish Solar Telescope) provides a unique opportunity to 
study these short-lived, propagating chromospheric waves in detail. Such a study is not possible, at present, 
for the coronal counterpart \citep{tom,erd3}.

\begin{acknowledgements}
Observations were obtained at the National Solar Observatory, operated by the 
Association of Universities for Research in Astronomy, Inc (AURA) under 
agreement with the National Science Foundation. 
RE acknowledges M. K\'eray for patient encouragement. The authors are also grateful to NSF, 
Hungary (OTKA, Ref. No. K83133). This work is supported by the UK Science 
and Technology Facilities Council (STFC), with DBJ particularly grateful for the 
award of an STFC post-doctoral fellowship. We thank the Air Force Office of Scientific Research, Air Force 
Material Command, USAF for sponsorship under grant number FA8655-09-13085.
\end{acknowledgements}

\end{document}